\documentclass[11pt]{article}
\usepackage{geometry}
\usepackage{graphicx}
\usepackage{authblk}
\usepackage{amsfonts}
\usepackage{latexsym,amssymb,amsmath}
\usepackage[boxed]{algorithm2e}
\renewcommand{\thealgocf}{}
\usepackage{comment}

\providecommand{\DontPrintSemicolon}{\dontprintsemicolon}

\newtheorem{theorem}{Theorem}
\newtheorem{remark}{Remark}
\newenvironment{proof}{\noindent{\bf Proof.}}{\hfill \qed \vskip 5pt}
\def\qed{\hfill\rule{2mm}{2mm}}
\begin{document}

\title{Linear Search with Terrain-Dependent Speeds\footnote{Research supported in part by NSERC Discovery grants}}
\author[1]{J. Czyzowicz}
\author[2]{E. Kranakis}
\author[3] {D. Krizanc}
\author[4]{L. Narayanan}
\author[4]{J.~Opatrny}
\author[5]{S. Shende}
\affil[1]{D\'{e}pt. d'Inf.,
Universit\'{e} du Qu\'{e}bec en
Outaouais, Gatineau, Canada}
\affil[2]{School of Comp. Science, Carleton University, Ottawa, 
Canada}
\affil[3]{Dept. of Math. and Comp. Science,
Wesleyan University, Middletown CT, USA}
\affil[4]{Dept. of Comp. Sc. and Soft. Eng., 
Concordia University, Montreal, QC,  Canada}
\affil[5]{Dept. of Comp. Science, Rutgers University, Camden, USA}
\renewcommand\Authands{ and }


\maketitle

\begin{abstract}

We revisit the \textit{linear search} problem where a robot, 
initially placed at the origin on an infinite line, tries to locate 
a stationary target placed at an unknown position on the line. 
Unlike previous studies, in which the robot travels along the line 
at a constant speed, we consider settings where the robot's speed 
can depend on the direction of travel along the line, or 
on the profile of the terrain, e.g. when the line is inclined, and 
the robot can accelerate. Our  objective is to design search algorithms 
that achieve good 
{\em competitive ratios} for the {\em time} spent 
by the robot to complete its search versus the time spent by an
omniscient robot that knows the location of the target. 

We consider several new robot mobility models in which
the {\em speed} of
the robot depends on the terrain.  These include 1) different constant
speeds for different directions, 2) speed with constant acceleration 
and/or variability depending on whether a certain segment has 
already been searched, 3) speed dependent on the incline 
of the terrain. We provide both upper and 
lower bounds on the competitive ratios of search algorithms for these models, and in 
many cases, we derive \textit{optimal} algorithms for the search 
time. 

\vspace{0.5cm}

\end{abstract}


\section{Introduction}

Searching and exploration 
are fundamental problems in the areas of robotics and autonomous 
mobile agents. The objective for searching is to find a target 
placed at an unknown location in the domain in a provably optimal 
manner. In the \textit{linear search} problem, 
the target is placed at a location on 
the infinite line unknown to the robot.
The robot moves with uniform speed, and the goal is to find the target in 
minimum
time. This problem was first proposed by Bellman~\cite{bellman1963optimal} and independently by Beck~\cite{beck1964linear}. 

Previous studies on the linear search problem generally assume that 
the robot moves with constant speed that is independent of the terrain.  In 
this paper, we study a generalization of the problem where the 
horizontal 
line may be replaced by a more complicated (and hence, more 
realistic) continuous \textit{linear terrain}.
Moreover, the speed of the robot may depend in various ways 
on the nature or profile of the terrain. The robot initiates the search for 
the unknown target on the terrain from a reference starting point 
(without loss of generality, the origin). In our models,  the robot 
can move with {\em different speeds} depending on its position on 
the terrain, its direction of movement, its exploration history etc. We also 
assume that the robot starts moving initially in the positive $x$-
direction, or more informally, moving to the \textit{right} (the 
\textit{leftward} movement is in the negative $x$-direction).

Consider the linear search problem with a single robot. 
Since the position of the target is unknown to the robot, the
robot cannot proceed indefinitely in just one direction and is 
forced to turn around and explore the terrain in the opposite
direction as well; this \textit{zig-zag} 
movement  is inevitable and must be repeated periodically.
The \textbf{canonical zig-zag search algorithm} is described below: note that the algorithm is parametrized by an infinite sequence of positive distances 
$X=\{x_k\}_{k\geq 1}$ from the origin that specifies the turning points. We refer to the sequence $X$ as the {\em strategy}. To ensure progress in searching 
along a given direction, each trip away from the origin must cover 
\textit{more} distance along the line than the previous trip in the 
same direction: this is formalized in the requirement that 
$x_k < x_{k+2}$ for all $k \geq 1$. 

\begin{algorithm}
\DontPrintSemicolon
\SetKwInOut{Input}{Input}
\Input{Infinite sequence of distances $X=\{x_1, x_2, \ldots\}$ with 
$0 < x_k < x_{k+2}$  for all $k \geq 1$}
\BlankLine
\For{$k \leftarrow ~1, ~2, ~\ldots$}{
	\If{$k$ is \textit{odd} (resp. \textit{even})}{
		move right (\textit{resp.} left) a distance of $x_k$ unless 
			the target is found enroute\;
		\If{target found}{
				\textbf{quit search}}
		Turn; then move left (\textit{resp.} right), return to 
origin\;}
} 
\BlankLine
\caption{\textbf{Zig-Zag Search}}
\end{algorithm}

A natural measure of the efficacy of the zig-zag search algorithm with strategy $X$, is how well it performs in competition with an omniscient adversary that knows the exact location of the target. 
Let $\sigma_X(d)$ be the ratio between the \textit{time} taken by the robot using the zig-zag strategy $X$ to reach an unknown target at distance $d$ from the origin versus the \textit{time} taken by the  adversary to proceed directly to the target.
Then, 
\[ \sigma_X \triangleq \sup_{d>1} ~\sigma_X(d) \]
denotes the \textbf{competitive ratio} of the algorithm. We denote the optimal competitive ratio by $\sigma^*$.

For strategies where $x_k=\alpha r^{k-1}$ for some constant $\alpha > 0$, we call $r$ the {\em expansion factor} of the strategy. Let $D$ denote the \textit{doubling strategy} that is a strategy with expansion factor $2$ (and $\alpha=1$). Thus, $D=\{1,~2,~2^2,\ldots\}$. When the robot moves with unit speed in both directions, it 
is well-known that the doubling strategy is optimal:  
 $\sigma^* = \sigma_D = 9$, see for example \cite{baezayates1993searching}.

\subsection{Our Results}

A natural point of departure from traditional unit-speed models
is to considering \textit{linear terrains} in which the speed of the robot \textit{depends} on the nature of the terrain or the environment.
Two kinds of models are considered:

\begin{figure}[htb]
\begin{center}
\includegraphics[scale=.5]{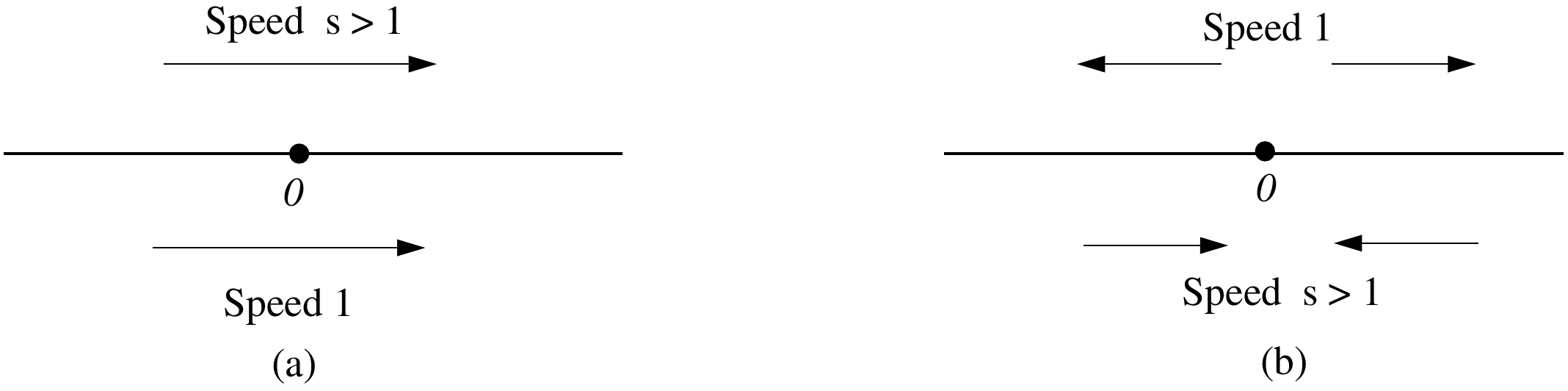}
\end{center}
\hspace*{-0.25in}
\caption{Two-speed models based on (a) absolute direction and (b) direction relative to origin}
\label{fig:linsearch_twospeed}
\end{figure}

\noindent
(1) \textbf{Two-speed models of linear search: } The robot can 
operate at two distinct constant speeds $1$ and $s>1$ in the following models.
\begin{itemize}
  \itemsep=1mm
\item The absolute direction or \textit{tailwind} model, viz. unit speed
  going left and tailwind speed $s>1$ going right (see Figure \ref{fig:linsearch_twospeed}-(a))
\item The direction relative to the origin or the \textit{beacon} model, viz. unit speed moving away from the origin and speed $s$ moving towards it (see Figure \ref{fig:linsearch_twospeed}-(b))
\item The \textit{exploration history} model, where the robot explores unknown regions slowly and deliberately with unit speed, but is able to search faster (with speed $s$) when it encounters a region already seen earlier in its search. 
\end{itemize} 

For the tailwind model, we analyze a \textit{time-based} zig-zag search strategy in Subsection~\ref{left or right: subsec}, which is provably better than the doubling strategy.  It turns out that the doubling zig-zag strategy  is \textit{optimal} for the beacon model; we prove this in Subsection~\ref{Moving Towards or Away from the Origin: subsec}. We also show in Subsection~\ref{Slower Speed when Exploring a New Section: subsec}
that the exploration history model admits an asymptotically optimal strategy, whose expansion factor depends on the speed $s$. 

\begin{figure}[htb]
\begin{center}
\includegraphics[scale=.55]{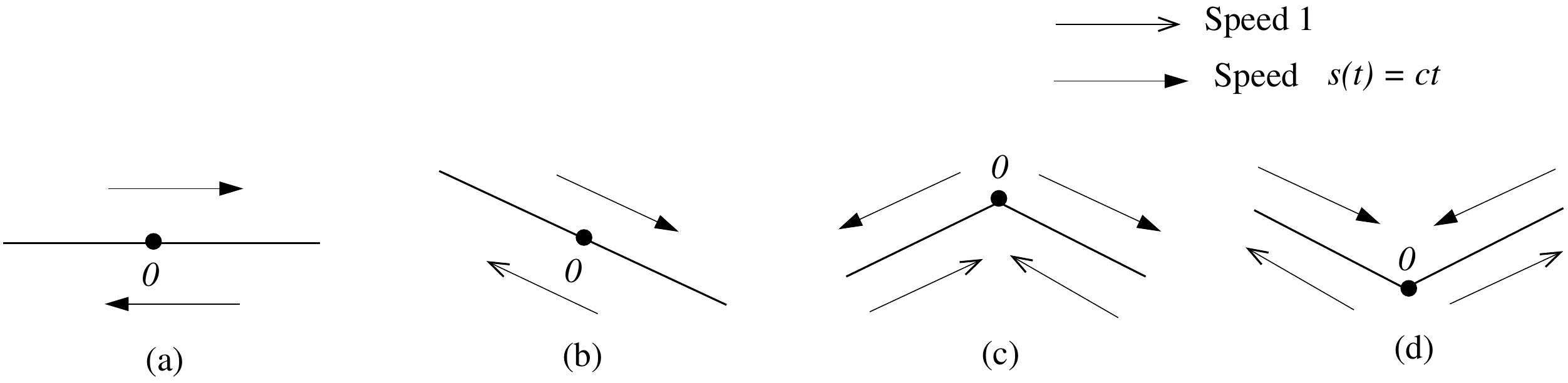}
\end{center}
\caption{Constant acceleration models: (a) Line, (b) Inclined line, (c) Hill, (d) Valley}
\label{fig:constant_accl}
\end{figure}

\noindent
(2) \textbf{Constant acceleration models for linear terrain search: } We first consider a linear search model with the property that whenever the robot starts from rest (i.e. either initially from the origin, or when it turns around in the zig-zag search), its speed increases at a constant rate $c$ until the next turn, i.e. at time $t$ after starting from rest, the robot's speed is given by $s(t)=ct$: see Figure \ref{fig:constant_accl}-(a). In  
Subsection~\ref{const-acc}, we show that for this model, ~$6.36 < \sigma^* < \sigma_D \approx 11.1$.

We then study search on \textit{inclined linear terrains}. The robot can operate in two modes where it is moving with unit speed \textit{going uphill} and with constant acceleration going \textit{downhill}. The different terrains include an inclined line, a symmetric hill with the hill-top at the origin, or a symmetric valley with the valley-bottom at the origin as shown in Figure \ref{fig:constant_accl}-(b), (c) and (d) respectively.  
Again, at time $t$ from rest, the robot's speed \textit{going downhill} is given by $s(t)=ct$. 
The increase in speed due to constant acceleration going downhill on a slope is a very natural manifestation of Newtonian physics: for example, we could interpret the constant ~$c$~ as the gravitational acceleration along the incline. 

We analyze the doubling zig-zag strategy and lower bounds on the optimal strategy for the inclined line, hill, and valley models in   Subsections~\ref{inclined1:subsec}, ~\ref{inclined2:subsec}, and ~\ref{inclined3:subsec} respectively. There are surprising differences in the nature of the results: while the competitive ratio of the doubling strategy is unbounded in the inclined line and hill models, we show that in the valley model, the competitive ratio is constant.

\subsection{Related Work}

Searching an environment or terrain with one or more searchers, 
possibly moving at different speeds, has the objective of localizing a 
hidden target in the minimum amount of time. Numerous variants of the 
search problem have been considered, e.g. with static or moving 
targets, multiple searchers with or without communication 
capabilities, and in environments that may not be fully known in 
advance.

The search problem has been extensively studied, e.g. see the survey by Benkoski \textit{et.al.} \cite{benkoski1991survey}; deterministic 
algorithms for optimal linear search \cite{baezayates1993searching}; 
incorporating a \textit{turn cost} when a robot changes direction 
during the search \cite{demaine2006online}; when bounds on the 
distance to the target are known in advance \cite{Bose13}; and for
moving targets or more general linear cost functions \cite{Bose16}. 
Other approaches include optimal randomized algorithms for the 
related \textit{cow-path problem} \cite{kao1996searching}, and 
stochastic and game theoretic investigations 
\cite{alpern2002theory,beck1973return}.

The search problem has also been studied in environments where search occurs in  graphs (see, e.g. \cite{FT08}) or along dynamically evolving links of
networks \cite{casteigts2011,kuhn2010}.  More recently, variants of
search using collections of \textit{collaborating} robots have been
investigated. The robots can employ either \textit{wireless} communication (at any
distance) or \textit{face-to-face} communication, where communication is
only possible among co-located robots. For example, the problem of \textit{evacuation} \cite{CGKNOV, CKKNOS} is essentially a search problem where search is completed only when the target is reached by the last robot.
Linear group search in the face-to-face communication model has also been studied with robots that either operate at the same speed or with a pair of robots having distinct maximal speeds \cite{Groupsearch, SIROCCO16}. Finally, a new direction of research seeks to analyze linear search with multiple robots where some fraction of the robots may exhibit either {\em crash 
faults} \cite{PODC16} or \textit{Byzantine faults} \cite{ISAAC16}.

\section{Two-speed Models of Linear Search}
\label{ConstantSpeeds: sec}

In this section we consider linear search problems where the robot can switch between two different constant speeds depending on its absolute direction of movement (the \textit{tailwind} model) or its direction of movement relative to the origin (the \textit{beacon} model).

\subsection{The Tailwind Model}
\label{left or right: subsec}

In this model, the robot moves at speed $s>1$ in the positive (right) direction and at unit speed in the negative (left) direction as depicted in Figure~\ref{fig:linsearch_twospeed}-(a).  
Observe that if we use 
the doubling strategy, the size of the explored segment expands by a factor of 2 in each iteration (i.e. between turns). However, the strategy favours the 
negative direction of the line  in the sense that {\em  it spends less
time exploring the positive direction of the line}
because the speed is higher when moving right. 

To account for this, we propose a different strategy, \textit{viz.} one that \textit{balances} the search time on \textit{both sides} of the origin. In other words, we expand the \textit{time} spent on each side of the origin, rather than the distance travelled as follows.
Fix two parameters $r > 1$ and
$\alpha > 0$. Then, our strategy is defined as 
sequence $X=\{x_1, ~x_2, \ldots\} = \{s, ~\alpha r, ~r^2 s, ~\alpha r^3, \ldots\}$, i.e. with 
$x_{2k-1} = r^{2k-2} s$ and $x_{2k} = \alpha r^{2k-1}$ for $k \geq 1$. 

Thus, strategy $X$ spends \textit{even powers} of $r$ time moving to the right from the origin, and $\alpha$ times \textit{odd powers} of $r$ time moving the left. 
In the next theorem we show how to select the parameters $\alpha , r$ so as to optimize the search time. In particular, we prove the following result.

\begin{theorem}
\label{bal:th}
Assume the robot has speed $s \geq 1$ when moving left to right and speed $1$ otherwise. 
For $\alpha, r$ such that 
$
\alpha = (1-s + \sqrt{(s-1)^2 + 4r^2 s})/(2r) 
\mbox{ and } 
r = \sqrt{ 2 + ({s+1})/{\sqrt{s}}  }
$ , and  $X=\{s, \alpha r, r^2s, \alpha r^3, \ldots \}$ : 
\begin{equation}
\label{par0:eq}
2 + 1/s \leq \sigma^* \leq \sigma_X \leq 1+ \frac{s+2\sqrt{s}+1}{s+\sqrt{s}+1} \cdot \frac{s+1}{2s} \cdot
\left( s+1 + \sqrt{(s-1)^2 + 8s + 4 \sqrt{s} (s+1)} 
\right) .
\end{equation}
\end{theorem}

\begin{proof} 
Let's look first at the lower bound. The robot must visit both points $+d$ and $-d$. If $-d$ is the first point to be visited by the robot then the adversary will place the target at $+d$. The resulting competitive ratio in this case is at least $(d+ 2d/s)/(d/s) = 2+s$. If $+d$ is the first point to be visited by the robot then the adversary will place the target at $-d$. The resulting competitive ratio in this case is at least $(d/s + 2d)/d = 2+ 1/s$. Since $s \geq 1$, the minimum of these two values is $2+ 1/s$, which proves that $\sigma^* \geq 2 + 1/s$.

Next we consider the upper bound given by the strategy $X$. 
There are two possibilities depending on whether the target is located at $+d$ or $-d$.: 
\paragraph{Case 1: The target is at $+d$.} Let $i$ be defined such that
$r^{2i}s < d \leq r^{2i+2}s$. The time $T$ it takes until the target is given by:
\begin{align*}
T&= r^0 + r^0 s + \alpha r^1 + \alpha r^1/s + \cdots +  r^{2i} + r^{2i} s + \alpha r^{2i+1} + \alpha r^{2i+1}/s + d/s \\
&= (1+s) (r^0 + r^2 + \cdots + r^{2i}) + \alpha r (1+1/s) (r^0 + r^2 + \cdots + r^{2i}) + d/s \\
&= (1 + s) (1 + \alpha r /s) \frac{(r^2)^{i+1} - 1}{r^2 -1} + d/s
<  (1 + s) (1 + \alpha r /s) \frac{(d/s)r^2}{r^2 -1} + d/s 
\end{align*}
and an upper bound on $\sigma_X(d)$ for this case (denoted $\sigma^+$) is obtained by dividing the last term above by $d/s$, namely 
\begin{equation}
\label{par1:eq}
\sigma^{+} := 1 +  (1 + s) (1 + \alpha r /s) \frac{r^2}{r^2 -1} .
\end{equation}

\paragraph{Case 2: The target is at $-d$.}  Let $i$ be defined such that $\alpha r^{2i+1} < d \leq \alpha r^{2i+3}$. This is similar to the previous case with the additional term $r^{2i+2} +r^{2i+2}s $ and replacing $d/s$ by $d$. The time $T$ it takes until the target is given by:
\begin{align*}
T&= (1 + s) (1 + \alpha r /s) \frac{(r^2)^{i+1} - 1}{r^2 -1} + r^{2i+2} (1+s)  + d \\
& <  r^{2i+2} (1+s) \left(  1 + \frac{1+ \alpha r /s }{r^2 -1}\right) + d
< \frac{rd}{\alpha} (1+s) \frac{r^2 + \alpha r /s}{r^2-1} + d
\end{align*}
and an upper bound on $\sigma_X(d)$ for this case (denoted $\sigma^-$)  is obtained by dividing the last term above by $d$, namely 
\begin{equation}
\label{par2:eq}
\sigma^{-} := 1 +  \frac{r^2}{r^2-1} (1 + s) (r /\alpha + 1 /s).
\end{equation}
We see that $\sigma^{+} = \sigma^{-}$ if and only if the following equation is satisfied
\begin{equation}
\label{par3:eq}
1+ \alpha r /s = r/\alpha + 1/s.
\end{equation}
If we multiply Equation~\eqref{par2:eq} by $\alpha s$ we derive the
equivalent quadratic (in $\alpha$) equation
$$r \alpha^2 +(s-1)\alpha - r s = 0$$ whose unique positive solution is 
$\alpha = (1-s + \sqrt{\Delta})/(2r)$ ,
where $\Delta = (s-1)^2 + 4r^2 s$.

Let $\sigma \triangleq \sigma_X(d)$ be the common value of $\sigma^+ , \sigma^{-}$ which is obtained for $\alpha = ({1-s + \sqrt{\Delta}})/(2r)$.

Observe from Equations~\eqref{par1:eq}~and~\eqref{par3:eq} that 
\begin{align*}
(\sigma -1) \frac{2s}{s+1}  
&= \frac{2s}{s+1} (s+1) (1 + \alpha r /s) \frac{r^2}{r^2 -1}
= 2 (s + \alpha r) \frac{r^2}{r^2 -1}
= (s+1 + \sqrt{\Delta}) \frac{r^2}{r^2 -1}=\\ 
&= (s+1 + \sqrt{\Delta}) \left( 1 + \frac{1}{r^2 -1} \right).
\end{align*}

Next, we minimize $\sigma$ as a function of the expansion factor $r$: it is straightforward to show that this is equivalent to solving the 
following equation of degree $4$ in the unknown $R$ where $R=r^2$.
\begin{align}
\label{final2:eq}
sR^4 - 6sR^3 + (9s -(s-1)^2) R^2 + (2 s^2 - 8s +2) R - (s-1)^2 = 0 .
\end{align}
Solving for $R$ and using $R=r^2$, we conclude that $r = \sqrt{2 + ({s+1})/{\sqrt{s}}}$
is the unique expansion factor which minimizes the competitive ratio $\sigma$ of our algorithm.
Substituting this value of $r$ into the formula for the competitive ratio in the right-hand side of Equation~\eqref{par1:eq}
yields the value
$$
1+ \frac{s+2\sqrt{s}+1}{s+\sqrt{s}+1} \cdot \frac{s+1}{2s} \cdot
\left( 
s+1 + \sqrt{(s-1)^2 + 8s + 4 \sqrt{s} (s+1)}
\right) ,
$$
which is exactly the right-hand side of Inequality~\eqref{par0:eq}.
This competes the proof of Theorem~\ref{bal:th}.
\end{proof}

\begin{remark}
Note that  as $s \to \infty$, the righthand side of Inequality~\eqref{par0:eq}  approaches $\frac32 + s + o(s)$.
\end{remark}

\subsection{The Beacon Model}
\label{Moving Towards or Away from the Origin: subsec}

In this model the robot moves with speed $1$ away from the origin and constant speed $s>1$ towards the origin of the line.
\begin{theorem} 
\label{tow1:th}
The doubling strategy is \textbf{optimal} for the beacon model, i.e.\begin{equation}
\label{awayor:eq}
\sigma^* = \sigma_D = 5+\frac{4}{s}.
\end{equation}
\end{theorem}

\begin{proof}
First we prove the upper bound.
Assume the robot executes Algorithm 1 with the doubling strategy,  and let the target be at distance $d$ from the origin. 
Let $k$ be such that $2^k < d \leq 2^{k+1}$. Since $2^k < d$, starting from the origin, by the $k$-th iteration of the algorithm the robot spends search time 
$
2^0 + 2^0/s + 2^1 + 2^1/s + \cdots + 2^k + 2^k/s = (2^{k+1} -1) (1+1/s)
$ 
and returns to the origin without having found the target. In the next turn, the robot again starts from the origin, spends time $2^{k+1} + 2^{k+1}/s$ and returns back to the origin, since the adversary could place the target to the other side of the origin. Hence the total time spent so far is
$
(2^{k+2} -1) (1+1/s)
$ 
Finally, since $d \leq 2^{k+1}$, in the last turn the robot finds the target in time $d$ 
It follows that 
\begin{align*}
\sigma_D & = 
\sup_{d>0}\frac{(2^{k+2} -1) (1+1/s)+d}{d}< \frac{4d (1+1/s) +d}{d}
= 5 + 4/s .
\end{align*}

 For the lower bound, we use a lower bounding technique \cite{car16} (itself based on 
\cite{baezayates1993searching}) used to obtain a lower bound of 9 for the unit speed model. Consider a deterministic strategy  $X= (x_1,x_2,\ldots)$ with $x_i > 0$, for all $1 \leq i < \infty$. We consider several cases depending on the position of the target.

Assume the target is between $x_k$ and $x_{k+2}$. Then the time it takes to find the target is 
\begin{align*}
& x_1 + x_1/s + x_2 + x_2/s + \cdots + x_k + x_k/s + x_{k+1} + x_{k+1} /s +  d 
= (1+1/s) \sum_{i=1}^{k+1} x_i + d .
\end{align*}
It follows that the competitive ratio is
\begin{align*}
\sigma_X &= \sup_k \sup_{d > x_k} \left\{ 1+ (1+1/s) \frac{\sum_{i=1}^{k+1} x_i }{d} \right\}
= \sup_k \left\{ 1+ (1+1/s) \frac{\sum_{i=1}^{k+1} x_i }{x_k} \right\}
\end{align*}
As a consequence it is easily seen that 
\begin{align*}
\sigma_X &\geq 2 + 1/s + (1+1/s) \frac{\sum_{i=1}^{k-1} x_i }{x_k} + (1+1/s) \frac{x_{k+1}}{x_k}
\end{align*}
and hence, 
$\sigma_X x_k \geq (2+1/s)x_k + (1+1/s) \sum_{i=1}^{k-1} x_i + (1+1/s) x_{k+1} $.
Separating $x_{k+1}$ from the last inequality we derive
\begin{align*}
x_{k+1} \leq \frac{\sigma_X - (2+1/s)}{1+1/s} x_k - \sum_{i=1}^{k-1} x_i.
\end{align*}
If we now put $\mu_0 = (\sigma_X - (2+1/s))/(1+1/s)$ and $\nu_0 =1$, the last inequality can be rewritten as
\begin{align}
\label{ind00}
x_{k+1} &\leq \mu_0 x_k - \nu_0 \sum_{i=1}^{k-1} x_i ,
\end{align}

At this point, we can use the technique suggested in \cite{baezayates1993searching,car16} as follows. 
Induction on the Inequality ~\eqref{ind00} can be used to construct
two infinite sequences of \textbf{positive} integers
$\{ \mu_{i}: i\geq 0 \}$ and   $\{\nu_i: i \geq 0 \}$ 
defined via a system of recurrences of the 
  form:
\begin{align}
\mu_{m+1} &= \label{rec1} \mu_0\mu_m - \nu_m\\
\nu_{m+1} &= \label{rec2} \nu_0\mu_m  + \nu_m.
\end{align}
for all $m \geq 0$.
The recurrences~\eqref{rec1},~\eqref{rec2} can be solved using 
difference equations
  that yield the characteristic polynomial 
$z^3 - \mu_0 z^2 + (\mu_0 +1) = 0$. This polynomial has $z=-1$ as
  one of its roots. Dividing by $z+1$ we obtain the polynomial  
\begin{align}
\label{rec4} 
z^2 - (\mu_0 +1) z + (\mu_0 +1) &=0
\end{align}
whose two roots are
$\rho_1, \rho_2 := \frac{\mu_0 +1 \pm \sqrt{D}}{2}$,
where $D := (\mu_0 -1)^2 - 4$ is the discriminant of the quadratic 
Equation~\eqref{rec4}. Note that $D < 0$ if and only if
${\mu}_0<3$. In turn, ${\mu}_0<3$ is equivalent to $\sigma_X$ being 
less than a certain constant $c$, since
$\sigma_X$ and ${\mu}_0$ are related (in the original proof of the 
lower bound, for instance, this yields $\sigma_X < c = 9$.)

On the other hand, it can be shown that $D$ being less than 0 
implies that for some $k\geq 1$, the value ${\mu}_k$ is negative. This 
is a contradiction since ${\mu}_{k}$ must be positive by construction for all $k$. Hence, $D$ must in fact be greater than or equal
to zero, and it follows that the CR $\sigma_X$ is greater than or equal 
to $c$.

By applying the above lower bound technique 
to the present case, we conclude that the roots of the resulting quadratic equation are conjugate complex numbers with non-zero imaginary parts iff $\mu_0 < 3$, which is equivalent to $\sigma_X < 5+4/s$.
This completes the proof of the lower bound of Theorem~\ref{tow1:th}.
\end{proof}

\begin{remark} Assume that the robot moves with speed  $1$ towards the origin, and speed $s$ away from the origin of the line. A proof similar to that of Theorem~\ref{tow1:th} shows that Algorithm 1 with the doubling strategy is optimal and its  competitive ratio is also $5+\frac{4}{s}$. Details are left to the reader.
\end{remark}

\subsection{The Exploration History Model}
\label{Slower Speed when Exploring a New Section: subsec}

We consider a robot that moves at speed $1$ when searching for the target, but the robot can move at speed $s>1$ when moving over a part of the line already explored. For example, the robot's attention 
to identifying the target limits the speed at which the robot can move.

\begin{theorem} 
\label{newsec:th}
Let $r= 1+\sqrt{2/(s+1)}$, and $X=(r^0,r^1,r^2,\ldots)$ be an expansion strategy. Then, with this strategy, the zig-zag algorithm's competitive ratio satisfies
\begin{equation}
\label{newsegment:eq}
2 + 1/s \leq \sigma^* \leq \sigma_X=2+\frac 1s \left(3+2\sqrt{2s+2} \right).
\end{equation}

\end{theorem}

\begin{proof} 
Consider first the lower bound. The robot must visit both points $+d$ and $-d$. Without loss of generality assume that $-d$ is the first point to be visited by the robot. Then the adversary will place the target at $+d$. Therefore the robot will traverse the segment $[-d, 0]$ once with speed at least $1$ to reach $-d$ from $0$ and a second time with speed $s$ on its way to $+d$ from $-d$. The resulting competitive ratio is at least $(2d+ d/s)/d = 2+ 1/s$.  This proves the lower bound.

Next we look at the upper bound. Consider a robot following a zig-zag strategy $X=(r^1, r^2, r^3, \ldots)$ and
that the first move of the robot is to the right with the target located at distance $d$ with $r^k <d \leq r^{k+2}$. The time needed by the robot to find the target is equal to
\begin{align*}
& r^1+r^1/s +r^2 +(r^2+r^1)/s 
+r^3-r^1 +(r^3+r^2)/s +r^4-r^2+\cdots \\
& ~+ (r^k +r^{k-2})/s + r^{k+1}-r^{k-1} + (r^{k+1}+r^{k})/s +d-r^k 
 =\left(1+\frac 1s \right)r^{k+1} +\frac 2s \sum_{i=1}^k\frac{r^i}s +d
\end{align*}

It follows that the competitive ratio of this strategy $\sigma_X$ is

\begin{align*}
\sigma_X 
&= \sup_{k\geq 1} \left( \left( \left(1+\frac 1s\right)r^{k+1} +\frac 2s \sum_{i=1}^k\frac{r^i}s +r^k\right)/r^k \right)
= \left(1+\frac 1s \right)r+ \frac{2r}{s(r-1)}+1 .
\end{align*}

To find the optimal value of $r$ we put the derivative  
$d\sigma_X/dr = 1+1/s-2/(s(r-1)^2)$ equal to $0$, which gives us that the competitive ratio is optimized for
$r=1+\sqrt{2/(s+1)}$
and for this $r$, we obtain
$\sigma_X=2+\frac 1s(3+2\sqrt{2s+2})$
\end{proof}

\begin{remark}
For example,  
if $s=1$ then $\sigma =9$, 
if $s=2$ then $\sigma\approx 5.95$, 
if $s=3$ then $\sigma \approx 4.88$, and
if $s=4$ then $\sigma \approx 4.33$.
Thus as $s \to \infty$ the value of $r$ approaches $1$ and the competitive ratio $\sigma $ as given in Theorem~\ref{newsec:th} approaches $2$. Therefore, the strategy is asymptotically optimal in $s$.
\end{remark}

\section{Searching with Constant Acceleration}
\label{ConstantAcceleration: sec}

In this framework, the robot exhibits constant acceleration $c >0$ in some part of the linear terrain when starting from rest. As is well known from Newtonian physics, at time $t$ after the robot accelerates from rest, it will be moving with speed $s=ct$ and would have covered a distance of $x(t)=ct^2/2$. Thus,  to cover distance $x$ we need time $\sqrt{2x/c}$.

\subsection{Constant acceleration in both directions}
\label{const-acc}
Here we assume that the constant acceleration applies in both directions throughout the entire terrain (see Figure~\ref{fig:constant_accl}-(a)).
\begin{theorem}
\label{tow2:th}
Assume the robot is searching with constant acceleration $c$ in either direction, starting from rest initially, as well as at turning points. Then:
\begin{equation}
\label{constacce:eq}
3 (\sqrt{2} + 1 / \sqrt{2}) \leq \sigma^* \leq \sigma_D \leq  \frac{2\sqrt{3} }{\sqrt{2}-1}  + \sqrt{3} + 1
\end{equation}
\end{theorem}

\begin{proof} 
First we consider the upper bound. 
Assume the robot executes the doubling strategy and let the target be at distance $d$ from the origin. 
 Let $k$ be such that $2^k < d \leq 2^{k+1}$.  There are two identical cases to consider depending on whether the target is to the left or right of the origin.

Let $b= \sqrt{2/c}$. Since $2^k < d$, starting from the origin, the robot spends search time 
\begin{align*}
T & = b + b\sqrt{2^0 + 2^1} + b\sqrt{2^1+2^2} + \cdots + b\sqrt{2^{k-1} + 2^k} +b \sqrt{2^k+2^{k+1}} + b\sqrt{2^{k+1} + d}\\
& = b + b\sqrt{3} (\sqrt{2^0} + \sqrt{2^1} + \sqrt{2^k} ) + b\sqrt{2^{k+1} + d}
 = b+ b \frac{\sqrt{3} }{\sqrt{2}-1} 2^{k/2+1} + b\sqrt{2^{k+1} + d}<\\
& < b+  b \frac{2\sqrt{3} }{\sqrt{2}-1} \sqrt{d} + b\sqrt{3d} \mbox{ (since $2^{k+1} < 2d$)}
 = b + b \left( \frac{2\sqrt{3} }{\sqrt{2}-1}  + \sqrt{3} \right) \sqrt{d}
\end{align*}
The target is to the left or right of the starting position and at distance $d$; therefore if we divide the above expression by $b \sqrt{d}$, it follows that 
$$ \sigma_D(d) \leq 2\sqrt{3}/(\sqrt{2}-1)  + \sqrt{3} + {1}/{\sqrt{d}}$$ 
Since the maximum value of $\sigma_D(d)$ is achieved for $d=1$, we substitute $d=1$ to obtain the upper bound on $\sigma_D$.
\vspace*{2mm}
\noindent
Next we consider the lower bound. We use a similar technique as in the proof of Theorem~\ref{tow1:th}. Consider a deterministic strategy  $X= (x_1,x_2,\ldots)$ with $x_i > 0$, for all $1 \leq i < \infty$. Assume $x_k < d \leq x_{k+2}$. In the sequel we use the abbreviation $b= \sqrt{2/c}$.
The time it takes to find the target placed at $d$ satisfies the following equation.
\begin{align*}
& b \sqrt{x_1} + b \sqrt{x_1 + x_2} + \cdots + b\sqrt{x_k + x_{k+1}} + b \sqrt{x_{k+1}  +  d}
=\\&= b \sqrt{x_1}  + b \sum_{i=1}^{k} \sqrt{x_i + x_{i+1}} + b \sqrt{x_{k+1}  +  d}
\end{align*}
Therefore the competitive ratio is given by  
\begin{align}
\notag
~~~\frac{  \sqrt{x_1}  + \sum_{i=1}^{k} \sqrt{x_i + x_{i+1}} + \sqrt{x_{k+1}  +  d} }{\sqrt{d}} 
&\geq \frac{\sqrt{x_1} + \frac{1}{\sqrt{2}} \sum_{i=1}^{k} (\sqrt{x_i}+ \sqrt{x_{i+1}}) + \frac{1}{\sqrt{2}} (\sqrt{x_{k+1}} + \sqrt{d})}{\sqrt{d}} \\
&\label{cr11} 
\geq 
\frac{\frac{2}{\sqrt{2}} \sum_{i=1}^{k+1} \sqrt{x_i} + \frac{1}{\sqrt{2}} \sqrt{d}}{\sqrt{d}} ,
\end{align}
where in the righthand side above we used the simple inequality $\sqrt{x+y} \geq \frac{1}{\sqrt{2}} (\sqrt{x} + \sqrt{y})$. Therefore the overall competitive ratio $\sigma$ satisfies
\begin{align*}
\sigma &\geq  \frac{\frac{2}{\sqrt{2}} \sum_{i=1}^{k+1} \sqrt{x_i} + \frac{1}{\sqrt{2}} \sqrt{d}}{\sqrt{d}}
=  \sqrt{2} \sum_{i=1}^{k-1} \frac{\sqrt{x_i}}{\sqrt{x_k}} + \frac{3}{\sqrt{2}}  + \sqrt{2} \frac{\sqrt{x_{k+1}}}{\sqrt{x_k}} .
\end{align*}
If we multiply out by $\sqrt{x_k}$ we conclude that 
$
\sqrt{x_{k+1}} \leq \frac{\sigma - 3/\sqrt{2}}{\sqrt{2} } \sqrt{x_k} -  \sum_{i=1}^{k-1} \sqrt{x_i} 
$.\\
If we set $\mu_0 := \frac{\sigma - 3/\sqrt{2}}{\sqrt{2} }$ then as usual we obtain the condition 
$
(\mu_0 -1)^2 < 4 \Leftrightarrow -1 < \mu_0 < 3
$
for the quadratic to have complex roots. Substituting for $\mu_0$ yields the lower bound $3 (\sqrt{2} + 1 / \sqrt{2})$. 
\end{proof}


\subsection{Moving on an inclined line}
\label{inclined1:subsec}

In this section we consider the situation where the robot has unit speed in one direction, but  in the other direction, due to the inclination of the line, the robot is subjected to a constant  acceleration $c$.

Consider a target at distance $d > 1$ from the origin. In the theorem below we show that the doubling strategy has unbounded competitive ratio. 
\begin{theorem}
\label{grav:th}
Assume the robot moves with acceleration $c$ in the positive direction, and constant speed $1$ in the negative direction using the doubling strategy.
 Then for any $d \geq 1$, 
$$\sqrt{2c}\sqrt{d}<\sigma_D(d)  \leq \sqrt{8 c} \cdot \sqrt{d}  + O(1)$$
Furthermore, $\sigma^* \geq \sup_{d>1} \min \{ 2 + \sqrt{2/(cd)}, \sqrt{2} + \sqrt{cd/2}  \}$.
\end{theorem}
\begin{proof} 
Let's look at the lower bound first. The robot must visit both points $+d$ and $-d$. Assume that $-d$ is the first point to be visited by the robot. Then the adversary will place the target at $+d$. Therefore the robot will traverse the segment $[-d, 0]$ in time at least $d$ and then move downhill a distance $2d$ to the target. Thus 
$$
\sigma^*\geq (d + \sqrt{{4d}/c})/\sqrt{{2d}/{c}} = \sqrt{ cd/2} + \sqrt{2} .
$$ 
Now assume that $+d$ is the first point to be visited by the robot. Then the adversary will place the target at $-d$. Therefore the robot will traverse the segment $[0, +d]$ in time at least $\sqrt{{2d}/{c}}$ and then move uphill a distance $2d$ to the target. Thus  
$$
\sigma^* \geq (\sqrt{{2d}/{c}} + 2d)/{d} = 2 + \sqrt{2/(cd)},
$$
which proves the lower bound.

Next we look at the upper bound. Assume the robot executes the doubling strategy and let the target be at distance $d$ from the origin. 
Let $k$ be such that $2^k < d \leq 2^{k+1}$ and $b=\sqrt{2/c}$.  There are two cases to consider depending on the parity of $k$.

\vspace*{-4mm}
\paragraph{The target is uphill and $k$ is even.}
Since $2^k < d$, starting from the origin, the robot spends search time 
\begin{align*}
& b + (2^0 + 2^1) + b\sqrt{2^1+2^2} + \cdots + (2^{k-1} + 2^k) +b \sqrt{2^k+2^{k+1}} + 2^{k+1} + d \\
& = 2^{k+2} -1 + b + b\sqrt{2^1+2^2}  + \cdots + b \sqrt{2^k+2^{k+1}} + d \\
& < 5d + b + b \sqrt{3} (\sqrt{2^0} + \sqrt{2^2} + \cdots + \sqrt{2^k}) 
= 5d + b \sqrt{3} (2^{k/2+1} -1)
\end{align*}
The target is uphill at distance $d$; therefore if we divide the above expression by $d$, we conclude that in this case  $\sigma_D(d) \leq 5+O(d^{-1/2})$.

\vspace*{-4mm}
\paragraph{The target is downhill and $k$ is odd.} Starting from the origin, the robot spends search time 
\begin{align*}
T=& 1 + b\sqrt{2^0+2^1} + 2^1+2^2 +\cdots + b\sqrt{2^{k-1} + 2^k} + 2^k+2^{k+1} + b\sqrt{2^{k+1} + d} \\
&= 2^{k+2} -1 + b\sqrt{2^0+2^1} + \cdots + b\sqrt{2^{k-1} + 2^k}  + \sqrt{2^{k+1} + d}\\
&=2^{k+2}-1+b\sqrt 3(2^{(k+1)/2}-1) + \sqrt{2^{k+1} + d}
\end{align*}
Since $2^k < d\leq 2^{k+1}$,
we have that
$$2d-1+  b \sqrt{3} (\sqrt{d}-1)<T < 4d + b \sqrt{3} \sqrt{2d} + b \sqrt{3d}$$
 The target is downhill at distance $d$; if we divide the above expression by $b\sqrt{d}$, we conclude that in this case 
 $$\frac{2d}{b\sqrt{d}}=\sqrt{2c}\sqrt{d}< \sigma_D(d) \leq   \frac{4}{b} \sqrt{d} + O(1) = \sqrt{8c}\cdot\sqrt{d} +O(1).
$$
\end{proof}


\subsection{Starting at the top of a hill}
\label{inclined2:subsec}

This model differs from the previous one by having the origin of the line located on the top of a hill. Thus, the speed of a robot increases when going downhill from
the origin.
Namely it travels with constant speed $1$ uphill but has a constant acceleration when going downhill. The main result here is that the competitive ratio of the optimal search algorithm is unbounded. Notice that if a robot has initial speed $1$ at the top of the hill then when going downhill with constant acceleration $c$ it has at time $t$ speed $1+ct$,  and to covers distance $x$ it needs time $(\sqrt{1+2cx}-1)/c$.   
\begin{theorem}
\label{grav1:th}
 Assume that the robot travels with constant acceleration $c$ away from the origin, and with unit speed towards the origin. Then 
 $\sigma_D(d) = \Theta (\sqrt{d})$ and this is optimal. 

\end{theorem}
\begin{proof} 
  The upper bound proof uses the main idea of the upper bound in Theorem~\ref{grav:th}. However, unlike in Theorem~\ref{grav:th}, the analysis of the algorithm is now symmetric. As before, let $k$ be such that $2^k < d \leq 2^{k+1}$ and $b=\sqrt{2/c}$.
Since $2^k <d$, starting from the origin, the robot spends search time 
\begin{align*}
  &b+1+(\sqrt{1+c2^2}-1)/c  + 2^1 +(\sqrt{1+c2^3}-1)/c +2^2+\\
  &+\cdots+(\sqrt{1+c2^{k+2}}-1)/c +2^{k+1} + (\sqrt{1+cd}-1)/c=\\
  &=b+\frac{1}{c}(-(k+1) + \sqrt{1+cd} +\sum_{i=2}^{k+2}\sqrt{1+c2^i}\;\;)+\sum_{i=0}^{k+1}2^i\\
   &<b+ \sqrt{1+cd}/c +b\sum_{i=3}^{k+3}\sqrt{2^i}+\sum_{i=0}^{k+1}2^i< b+2^{k+2} + b2^{(k+4)/2}+ \sqrt{1 + cd}/c \\
& < b+4d +  b (1+4\sqrt{d}) + \sqrt{1 + cd}/c \\
\end{align*}
 The target is downhill at distance $d$; if we divide the above expression by $b\sqrt{d}$, we get that the competitive ratio in this case is at most 
 $$
 \frac{b+4d + b(1+4\sqrt{d})+  \sqrt{1+cd}/c}{b\sqrt{d}} = \frac{4}{b} \sqrt{d} + O(1).
 $$

 \noindent To see the  lower bound, observe that the robot must visit both points $+d$ and $-d$. Assume that $-d$ is the first point to be visited by the robot. Then the adversary will place the target at $+d$. Therefore the robot will traverse the segment $[-d, 0]$ in time at least $b \sqrt{d}$. To get to $+d$, the robot needs time at least $d$ to get to the origin and another $(\sqrt{1+2cd}-1)/c=\sqrt{1/c^2+b^2d}$ to reach the target.
The omniscient optimal algorithm needs time $b \sqrt{d}$. Thus, for any strategy $X$, 
$$
\sigma_X(d) \geq \frac{b\sqrt{d} + d +\sqrt{1/c^2+b^2d}}{b \sqrt{d}} = 1 + \sqrt{d}/b +\sqrt{\frac{1}{c^2b^2d}+1}.
$$ 
\end{proof}

\subsection{Starting at the bottom of a valley}
\label{inclined3:subsec}

An interesting situation occurs if we reverse the speeds, i.e., the origin is located at the bottom of a valley and thus we have constant acceleration when moving towards the origin, but the robot moves at unit speed away from the origin (see~Figure~\ref{fig:constant_accl}).  
In this case can prove the following theorem: 
\begin{theorem}
\label{grav2:th}
 Assume that the robot travels with constant acceleration $c$ towards the origin, and with unit speed away from the origin. Then for any $d \geq 1$:
 $$\sigma_D(d) \leq 5+ O(d^{-1/2})$$
 Furthermore, $\sigma^* \geq 5$.
\end{theorem}
\begin{proof} As before, the upper bound proof uses the main idea of the upper bound in Theorem~\ref{grav:th}. However, unlike Theorem~\ref{grav:th}, the analysis of the algorithm is now symmetric. Let $k$ be such that $2^k < d \leq 2^{k+1}$ and $b=\sqrt{2/c}$. Since $2^k < d$, starting from the origin, the robot spends search time 
\begin{align*}
& 1 + b \sqrt{2^0} + 2^1 + b \sqrt{2^1}  + \cdots + 2^k + b \sqrt{2^k} + 2^{k+1} + b \sqrt{2^{k+1}} +d   \\
& = 2^{k+2} -1 + b \left( \sqrt{2^0} + \sqrt{2^1} + \cdots +  \sqrt{2^k} + \sqrt{2^{k+1}} \right) + d \\
& < 4d + b\frac{(\sqrt{2})^{k+2} - 1}{\sqrt{2} -1} + d
 < 5d + \frac{2b}{\sqrt{2}-1} \sqrt{d} 
\end{align*}
 The target is uphill at distance $d$; if we divide the above expression by $d$, we get that the competitive ratio in this case is at most 
 $$
\frac{5d + \frac{2b}{\sqrt{2}-1} \sqrt{d}}{d} =  5 + O(d^{-1/2}).
$$

Next we look at the lower bound.
Consider a deterministic strategy  $X= (x_1,x_2,\ldots)$ with $x_k > 0$, for all $1 \leq k < \infty$. 
Because of the symmetry of the problem, we may assume that the target is to the right of the origin  between $x_k$ and $x_{k+2}$ and $k$ is odd. Then the time it takes to find the target is 
\begin{align*}
& x_1 + b \sqrt{x_1} + x_2 + b \sqrt{x_2} + \cdots + x_k  + b \sqrt{x_k} + x_{k+1} +  b \sqrt{x_{k+1}}  + d 
= \sum_{i=1}^{k+1} x_i + b  \sum_{i=1}^{k+1} \sqrt{x_i} + d 
\end{align*}
Therefore the competitive ratio satisfies 
\begin{align*}
\sigma \geq \frac{\sum_{i=1}^{k+1} x_i + d}{d} \geq 1 + \sum_{i=1}^{k+1} \frac{x_i}{x_k} 
= 2 + \frac{x_{k+1}}{x_k} + \sum_{i=1}^{k-1} \frac{x_i}{x_k} .
\end{align*}
In turn, this gives rise to the following recurrence
\begin{align}
\label{crbottom}
x_{k+1} \leq (\sigma -2) x_k + \sum_{i=1}^{k-1} x_i .
\end{align}
Note that Inequality~\eqref{crbottom} is exactly of the form displayed in Recurrence~\eqref{ind00} with $\mu_0 = \sigma -2$ and $\nu_0 = 1$.  Moreover the same proof technique yields easily that the competitive ratio is at least $5$. 
\end{proof}

\section{Discussion}

In this paper we have considered and analyzed several zig-zag strategies for 
search on a linear terrain for cases when the speed of the robot is not constant. Our work provides an initial step for the study of a robot searching  terrains of different profiles for a target placed at an unknown location. We study  two kinds of models of speed: two-speed models, and constant acceleration models. An interesting observation is in our two-speed models, as in the traditional one-speed model, the performance of the doubling algorithm vis-a-vis an omniscient optimal algorithm gets worse as $d$ (the distance of the target to the initial location) increases and converges to some maximum value as
$d \rightarrow \infty$. However, in the constant acceleration models that we studied, either the competitive ratio is unbounded, or the performance of the doubling algorithm improves vis-a-vis an omniscient optimal algorithm as $d$ increases.

\bibliographystyle{plain}

\bibliography{refs}

\end{document}